# Solvent-Free High-Temperature Capillary Stamping of Stimuli-Responsive Polymers: Wettability Management by Orthogonal Substrate Functionalization

Fatih Alarslan,[$] Hanna Hübner,[&,§] Jonas Klein,[$] Karsten Küpper,[%] Joachim Wollschläger,[%] Markus Haase,[$] Markus Gallei,[&,§] Martin Steinhart[$,*]

[$] School of Biology and Chemistry and Center of Cellular Nanoanalytics, Universät Osnabrück, Barbarastraße 7, D-49076 Osnabrück, Germany

[&] Polymer Chemistry, Saarland University, Campus C4 2, 66123 Saarbrücken, Germany

[§] Saarland Center for Energy Materials and Sustainability, Campus C4 2, D-66123 Saarbrücken, Germany

[%] Department of Physics, Universität Osnabrück, Barbarastraße 7, D-49076 Osnabrück, Germany

**Abstract**

The wettability of surfaces determines their antifouling, antifogging, anti-icing and self-cleaning properties as well as their usability for sensing, oil-water separation, water collection and water purification. Solvent-free high-temperature capillary stamping of stimuli-responsive polymers yielding arrays of stimuli-responsive polymer microdots on differently modified substrates enables the flexible generation of switchable surfaces with different water contact angles (WCAs). Potential problems associated with the deposition of polymer solutions, such as the handling of volatile organic solvents, phase separation induced by solvent evaporation and capillarity-driven flow processes, are circumvented. We used composite stamps with topographically patterned contact surfaces consisting of metallic nickel cores and porous $MnO_2$ coatings taking up the stimuli-responsive polymers. The short transport ways from the $MnO_2$ contact layers to the counterpart substrates enabled the stamping of polymer melts containing components impeding flow, such as carbon nanotubes (CNTs). Thus-obtained arrays of polymer-CNT hybrid microdots prevent problems associated with continuous coatings including delamination and crack propagation. Moreover, the range within which the properties of the stamped stimuli-responsive polymer microdots are switchable can be tuned by orthogonal substrate modification. As example, we stamped hybrid microdots consisting of poly(2-(methacryloyloxy)ethyl ferrocenecarboxylate) (PFcMA) and CNTs onto indium tin oxide (ITO) substrates. Coating the ITO substrates with a PEO-terminated silane shifted the WCAs obtained by switching the PFcMA between its oxidized and reduced states by nearly 50°.

# 1. Introduction

The additive lithographic deposition of stimuli-responsive polymers[1] is frequently applied to produce functionalized surfaces. Classical additive lithographic methods deployed to this end include inkjet and aerosol jet printing[2] as well as soft lithography[3-6] and polymer pen lithography.[7-9] Commonly, solutions of the functional polymers in volatile organic solvents are lithographically deposited. The target patterns of the deposited functional polymers are obtained only after solvent evaporation. During the evaporation of volatile organic solvents hardly controllable structure formation processes, such as liquid-liquid phase separation[10] and flow processes related to, for example, capillarity[11, 12] may occur. Additional drying steps to remove residual solvent from as-deposited patterns may damage the latter. Moreover, the presence of volatile organic solvents reduces the amounts of the non-volatile functional polymeric components transferred per stamping step. If, on the other hand, functional polymers are transferred to substrates by ballistic methods, such as laser-induced forward transfer,[13] securing sufficient adhesion might be challenging. Melt processing of polymers by extrusion or electrospinning[14] is well established. Thus, the additive lithographic transfer of polymer melts in the absence of organic solvents might be a useful methodological extension of microcontact printing. In this way, polymers hardly soluble in solvents compatible with microcontact printing may be stamped also.

Wettability management is crucial to customize the antifouling, antifogging, anti-icing and self-cleaning properties of surfaces. Surfaces with tailored wettability may be used for applications like sensing, oil-water separation, water collection and water purification. Thus, the tailoring of the wettability of surfaces by additive surface manufacturing has attracted significant interest.[15-19] Surfaces with switchable wettability and, therefore, amplified functional versatility are accessible by functionalization with stimuli-responsive polymers.[20-22] Stimuli-responsive polymers may be chemically grafted onto substrates. However, non-reactive deposition significantly extends the range of deployable stimuli-responsive polymers and of counterpart substrates to be functionalized. Patterns of discrete stimuli-responsive polymer microdots may have additional advantages over continuous coatings, such as suppression of crack propagation and delamination. Moreover, the substrate properties may be customized by orthogonal functionalization of the substrate areas between the microdots.[23, 24]

Here, we show that solvent-free high-temperature capillary stamping is a potential manufacturing platform yielding tailored surfaces with switchable wettability by additive lithographic high-temperature deposition of stimuli-responsive polymers – optionally mixed with solid additives. As example, we performed high-temperature capillary stamping of melts of the stimuli-responsive ferrocene-containing polymer poly(2-(methacryloyloxy)ethyl ferrocenecarboxylate) (PFcMA)[25-27] blended with multiwalled carbon nanotubes (CNTs). We thus stamped arrays of PFcMA-CNT hybrid microdots onto conductive indium tin oxide (ITO) substrates. Electrochemical switching reversibly transformed the PFcMA-CNT hybrid microdots from a high-WCA state, in which the PFcMA is reduced, to a low-WCA state, in which the PFcMA is partially oxidized (the ferrocene units are positively charged), and *vice versa*. While the intrinsic conductivity of ITO is a prerequisite for the electrochemical switching of the PFcMA oxidation state, orthogonal functionalization of the ITO surface areas around the PFcMA-CNT hybrid microdots shifted the WCAs obtained for the two PFcMA oxidation states by nearly 50°.

Stamps exhibiting continuous pore systems[28-32] have been used for additive substrate patterning in automatized configurations[33] and even under the extreme conditions of solvothermal syntheses.[24] However, percolative transport of viscous mixtures containing polymers and additional components impeding flow, such as CNTs, through porous stamps might be too slow for efficient stamping procedures. Therefore, we devised Ni/MnO$_2$ composite stamps for high-temperature capillary stamping by adapting a composite design involving the functionalization of dip pens or PDMS stamps with porous coatings for the deposition of proteins,[34, 35] viruses[36] and bacterial cells.[37]

## 2. Materials and Methods
### 2.1. Chemicals and materials
Lithium perchlorate (LiClO$_4$), manganese sulfate (MnSO$_4$), nickel sulfate (NiSO$_4$), sodium acetate (C$_2$H$_3$NaO$_2$), sodium chloride (NaCl), boric acid (H$_3$BO$_3$), sodium dodecyl sulfate (NaC$_{12}$H$_{25}$SO$_4$; SDS), anisole (C$_7$H$_8$O), *tert*-butyl α-bromoisobutyrate (tBbib; C$_8$H$_{15}$BrO$_2$), copper (I) bromide (CuBr), pentamethyldiethylentriamine (PMDETA; C$_9$H$_{23}$N$_3$) and indium tin oxide (ITO) substrates [(In$_2$O$_3$)$_{0.9}$ • (SnO$_2$)$_{0.1}$ with a resistance of 8-12 Ω sq$^{-1}$ and a thickness of 1200–1600 Å] were purchased from Sigma-Aldrich. 3-[Methoxy(polyethyleneoxy)$_{6-9}$]propyltrichlorosilane CH$_3$O(C$_2$H$_4$O)$_{6-9}$(CH$_2$)$_3$Cl$_3$Si (PEO-silane) was purchased from Gelest. Multiwalled CNTs (NC7000) were purchased from Nanocyl. Macroporous silicon[38, 39] with macropores 1 μm in diameter and a depth of 730 nm arranged in a hexagonal array with a lattice constant of 1.5 μm was purchased from Smart Membranes (Halle). Pt wires were purchased from Rettberg Glasapparatebau and Ag/AgCl electrodes (c(KCl): 3 mol L$^{-1}$) from Metrohm. Polyimide tape (heat resistant adhesive 1810-DS tape) was purchased from M&S Lehner GmbH. If not stated otherwise, chemicals and materials were used as received.

## 2.2. Synthesis and characterization of PFcMA

1000 mg FcMA (4.35 mmol) and 4,7 µL tBbib (0.025 mmol) were dissolved in 5 mL anisole. The solution was degassed by 3 freeze-pump-thaw cycles and heated to 90 °C. After the quick addition of 375 µL of a CuPMDETABr solution (0.2 mol/L in anisole) the reaction mixture was stirred for 5 h. The polymer was then precipitated in a 10-fold excess of methanol, filtered, washed and dried in vacuo (SEC: $M_n$ = 13400 g/mol, $M_w$ = 21300 g/mol, $D$ = 1.58). The NMR spectrum of the obtained PFcMA was acquired at 400 MHz using a Bruker Avance II 400 spectrometer and processed with the software MestReNova (Mestrelab Research). The chemical shifts were referenced relative to the chemical shift originating from the used deuterated solvent (Figure S1). Size-exclusion chromatography (SEC) was carried out using an Agilent 1260 Infinity II setup with an SDV column set (1000 Å, 5000 Å, 6000 Å) from Polymer Standard Service (PSS, Mainz, Germany) and a PSS Security2 RI/UV detector. THF was used as the eluent at a flow rate of 1 mL/min and calibration was carried out using polystyrene standards (PSS, Mainz, Germany).

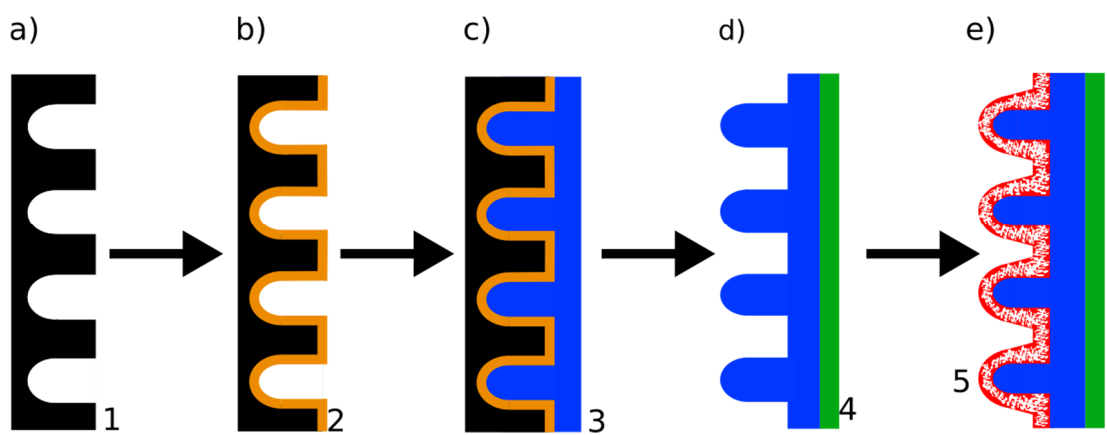

**Figure 1.** Preparation of Ni/MnO$_2$ composite stamps. a) Cross-sectional view of macroporous silicon (1; black) used as master template. b) Macroporous silicon coated with a thin gold layer (2; orange). c) Electrodeposition of nickel (3; blue) onto the gold-coated macroporous silicon. d) A heat-resistant adhesive polyimide tape (4, green) is glued on the flat exposed nickel surface and the nickel film supported by the polyimide tape is detached from the gold-coated macroporous silicon. e) A porous MnO$_2$ coating (5; red) is electrodeposited onto the topographically patterned surface of the nickel film initially in contact with the cold-coated macroporous silicon.

## 2.3. Preparation of Ni/MnO₂ composite stamps

Pieces of macroporous silicon[38, 39] extending 1 x 1 cm containing macropores with a diameter of 1 µm and a depth of 730 nm arranged in a hexagonal array with a lattice constant of 1.5 µm (Figure 1a) were sputtered 3 times with gold for 15 s using a JEOL JFC 1200 fine coater to generate an electrically conductive surface (Figure 1b).

Electrodeposition of nickel was performed in 20 mL of an aqueous solution containing 1 mol/L NiSO$_4$, 0.1 mol/L NaCl, 0.4 mol/L H$_3$BO$_3$ and 0.15 mol/L SDS.[40] A nickel plate was used as counter electrode and the gold-coated macroporous Si template as working electrode. The deposition was conducted with a current of 1 mA/cm$^2$ for 600 s. The formed nickel film covering the gold-coated macroporous silicon (Figure 1c) was rinsed with water and dried. Heat-resistant adhesive polyimide tape was glued onto the smooth exposed surface of the nickel film, which were then detached from the silicon master templates. Thus, we obtained nickel films with a thickness of 2 µm, which were topographically structured with hexagonal arrays of rods having a lattice constant of 1.5 µm at their exposed surface opposite of the adhesive polyimide tape (Figures 1d and S2). The rods were replicas of the macropores of the macroporous Si and had round tips, a height of 730 nm and base diameters of 1 µm. The adhesive polyimide tape mechanically stabilized the unsupported nickel film with an area of 1x1 cm$^2$. In the next step, the topographically structured exposed surface of the nickel film was coated with a porous MnO$_2$ layer with a thickness of ~200 nm (Figure 1e) by electrodeposition using a three-electrode system with a platinum counter electrode and an Ag/AgCl reference electrode. The electrodeposition was carried out for 90 s in 20 mL of an aqueous electrolyte solution containing 0.33 mol/L MnSO$_4$ and 0.33 mol/L sodium acetate with a current of 0.5 mA/cm$^2$ following procedures reported elsewhere.[41] All electrodeposition steps were performed with a potentiostat Interface 1000 (Gamry).

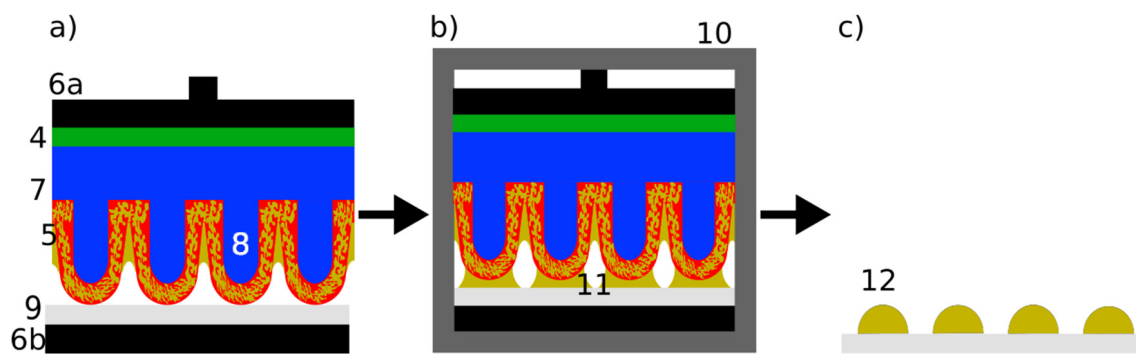

**Figure 2.** Solvent-free high-temperature capillary stamping of PFcMA-CNT hybrid microdots. a) Polyimide tape (**4**) of a Ni/MnO$_2$ composite stamp (**7**) is glued onto upper part (**6a**) of stamping device (**6**). ITO counterpart substrate (**9**) is placed on lower part (**6b**) of stamping device (**6**). The contact elements (**8**) of Ni/MnO$_2$ composite stamp (**7**) covered with MnO$_2$ layer (**5**) containing PFcMA-CNT mixture are brought into contact with ITO counterpart substrate (**9**). b) After insertion of stamping device (**6**) into vacuum furnace (**10**) and heating under Ar flow, the PFcMA-CNT mixture forms liquid bridges (**11**) between the contact elements (**8**) and ITO substrate (**9**). c) After detachment of Ni/MnO$_2$ composite stamp (**7**) and cooling to room temperature, ITO substrate (**9**) is modified with submicron PFcMA-CNT hybrid microdots (**12**).

## 2.4. High temperature capillary stamping

10 µL of a solution of 10 mg PFcMA and 0.2 mg CNTs in 10 mL toluene was pipetted onto the topographically patterned $MnO_2$ surface of the Ni/$MnO_2$ composite stamps in such a way that the whole stamp area extending 1 $cm^2$ was covered with a PFcMA-CNT film. After evaporation of the toluene under ambient conditions the PFcMA-CNT mixture was annealed at 190°C for 10 h (overnight) under a vacuum to remove residual solvent. The applied heating and cooling rates amounted to 20 K/min. For high-temperature capillary stamping we adapted a procedure reported elswhere[24] using a home-made stamping device (Figure S3). The flat nickel surface of the infiltrated Ni/$MnO_2$ composite stamps was glued onto the upper part of a stamping device with heat-resistant double-sided adhesive tape. ITO counterpart substrates to be patterned were located on the lower part of the stamping device. Prior to high-temperature capillary stamping, some of the ITO counterpart substrates had been surface-modified with PEO-silane by oxygen plasma treatment for 10 minutes using a Diener Femto plasma cleaner followed by chemical vapor deposition of PEO-silane at 110 °C for 12 h following procedures reported elsewhere.[42] The stamping device was then assembled by placing its upper part onto its lower part (Figure 2a). Thus, a contact pressure of 3.9 kN/$m^2$ was exerted during the stamping process. The assembled stamping device was then transferred into a home-made precision vacuum furnace (Figure 2b) and heated to 190°C at a rate of 20 K/min. The contact time at the target temperature of 190°C amounted to 1 min. The upper part of the stamping device was then removed and the lower part with the counterpart substrate was cooled to room temperature at a rate of -20 K/min. All high-temperature steps were carried out under argon atmosphere. After cooling to room temperature, the ITO substrates patterned with PFcMA-CNT hybrid microdot arrays (Figure 2c) could be picked up from the lower parts of the stamping device. To check whether the stamping procedure had an impact on the molecular weight distribution of the PFcMA, we stamped molten PFcMA without CNTs as described above on ITO substrates, scraped them off and used them for SEC measurements as described above.

## 2.5. Electrochemical analyses and treatments of PFcMA-CNT hybrid microdot arrays

For cyclic voltammetry we used ITO substrates with an area of 1 $cm^2$ functionalized with PFcMA-CNT hybrid microdots. The cyclic voltammetry measurements were performed in 20 mL of an aqueous electrolyte solution containing 0.1 mol/L $LiClO_4$ using a three-electrode setup with a platinum wire counter electrode and an Ag/AgCl reference electrode. The measurements were carried out at a scan rate of 20 mV/s. The oxidations and reductions of the PFcMA-CNT hybrid microdots for the wettability investigations were also carried out electrochemically. Oxidation and reduction steps were performed by applying a constant potential of 1 or 0 vs. Ag/AgCl, respectively, for 30 min in the setup described above.

## 2.6. Wettability tests

Apparent water contact angles (WCAs)[43] were measured by the sessile drop method with a DSA100 drop shape analyzer at 22 °C and a relative humidity of 23 %. To this end, we deposited droplets of deionized water with an initial volume of 2 µL on the sample surfaces. All data points represent the average of 3 measurements at different sample positions.

## 2.7. Microscopic characterization of PFcMA-CNT hybrid microdot arrays

For imaging by scanning electron microscopy (SEM), the samples were dried overnight at 40 °C in air and then sputter-coated 2-3 times for 15 s with platinum/iridium alloy using an EMITECH K575X sputter coater. SEM images were taken with a Zeiss Auriga device equipped with a field emission cathode and a Gemini column at a working distance of 5 mm applying an acceleration voltage of 3 kV. For image detection an InLens detector was used. Atomic force microscopy (AFM) topography images were recorded with an NTEGRA microscope (NT-MDT) in the tapping mode using HQ:NSC16/AL BS cantilevers from µmasch with a resonance frequency of 170 - 210 kHz and a force constant of 30 - 70 N/m.

## 2.8. Surface analysis by X-ray photoelectron spectroscopy

X-ray photoelectron spectroscopy measurements were carried out under ultra-high vacuum using an ESCA system Phi 5000 VersaProbe III with a base pressure of $1 \cdot 10^{-9}$ mbar equipped with a monochromatized aluminum anode (Kα = 1486.6eV) and a 32-channel electrostatic hemispherical electronic analyzer. An ion gun and an electron gun were used to prevent sample charging. A take-off angle of 45° was used. The XP spectra were calibrated using the carbon C 1s peak at 284.5 eV.[44] Lorentz fits of the XP spectra were obtained using the software Origin.

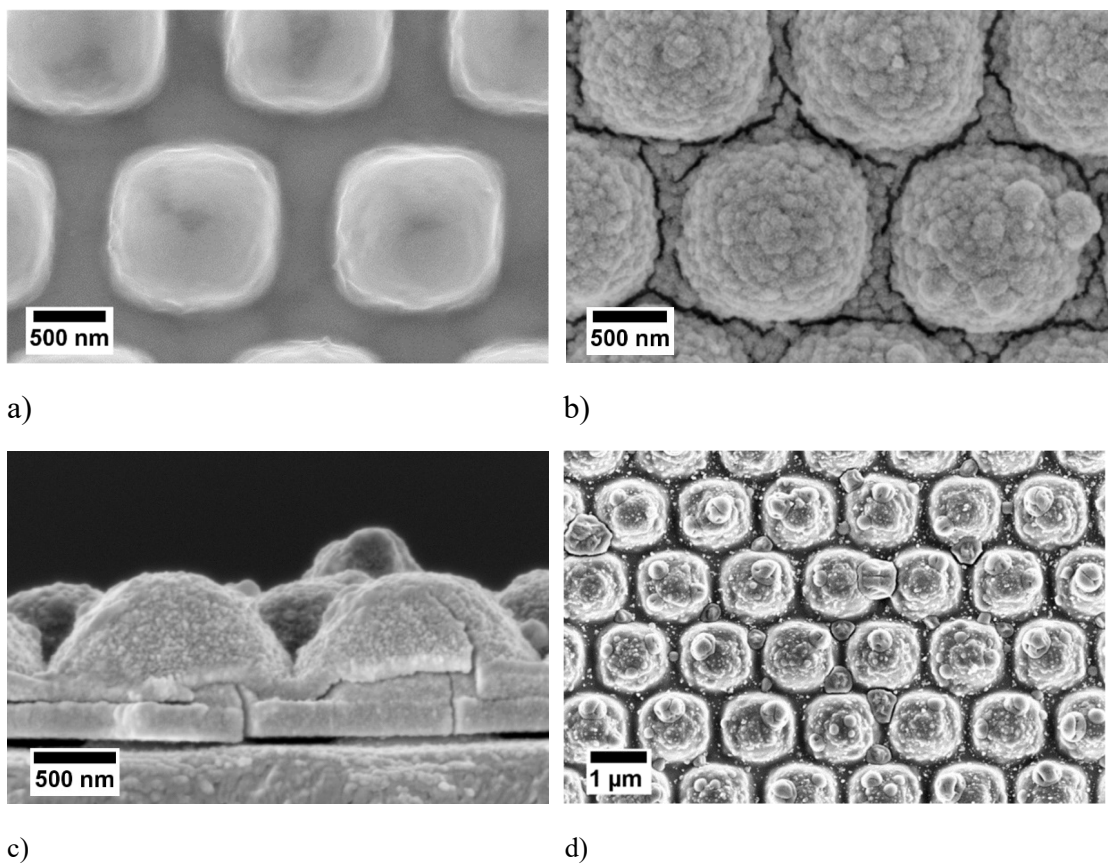

**Figure 3.** SEM images of Ni/MnO$_2$ composite stamps. a) Topographically patterned nickel layer prior to electrodeposition of MnO$_2$. b) Topographically patterned surface and c) cross section of a Ni/MnO$_2$ composite stamp. d) SEM image of the contact surface of a Ni/MnO$_2$ composite stamp after high-temperature capillary stamping.

## 3. Results and discussion
### 3.1. High-temperature capillary stamping

The Ni/MnO$_2$ composite stamps combined sufficient mechanical stability, stability at high temperatures, good heat conductivity and the capability of taking up sufficient amounts of the polymer melts to be transferred. The composite design enabled the deposition of mixtures containing components not able to flow, such as mixtures of molten polymers and CNTs, because the transport paths were shorter than in fully porous stamps. The Ni/MnO$_2$ composite stamps devised here with an area of 1 cm$^2$ contained nickel layers prepared by electrodeposition of nickel into macroporous silicon templates[38, 39] adapting procedures reported elsewhere (Figure 1a-c).[40] The surface of the ~2 µm thick nickel layers in contact with the macroporous silicon was patterned with arrays of nickel rods, which were negative replicas of the macropores of the macroporous silicon templates. Thus, the nickel rods arranged in hexagonal arrays with a nearest-neighbor distance of 1.5 µm had a diameter of 1 µm, a height of 730 nm and hemispherical tips (Figures 1d and 3a). SEM investigations revealed that no structural defects emerged from the detachment of the Ni layers from the macroporous Si templates (Figure S2). Nickel was selected because of its thermal conductivity and its mechanical properties.[45] These properties were combined with those of a ~200 nm thick porous $MnO_2$ layer with pore diameters of ~20 nm (Figures 1e and 3b,c) electrodeposited onto the topographically patterned nickel surface by adapting procedures reported elsewhere.[41] As a result, Ni/$MnO_2$ composite stamps having a topographically patterned porous $MnO_2$ contact surface were obtained. The topographic patterns of the contact surfaces consisted of arrays of rod-like contact elements with a height of ~730 nm and a diameter of 1.2 µm arranged in a hexagonal lattice with a lattice constant of 1.5 µm (Figure 3b,c). Using the Ni/$MnO_2$ composite stamps, we stamped PFcMA melts containing CNTs onto either pristine ITO substrates or ITO substrates modified with PEO-silane at 190°C under argon atmosphere. The contact time amounted to ~9.5 min including the heating ramp and a dwell time of 1 min at the target temperature. Liquid bridges consisting of PFcMA melts mixed with CNTs connecting the contact elements of the Ni/$MnO_2$ composite stamps and the surfaces of the ITO counterpart substrates were formed (Figure 2a,b).

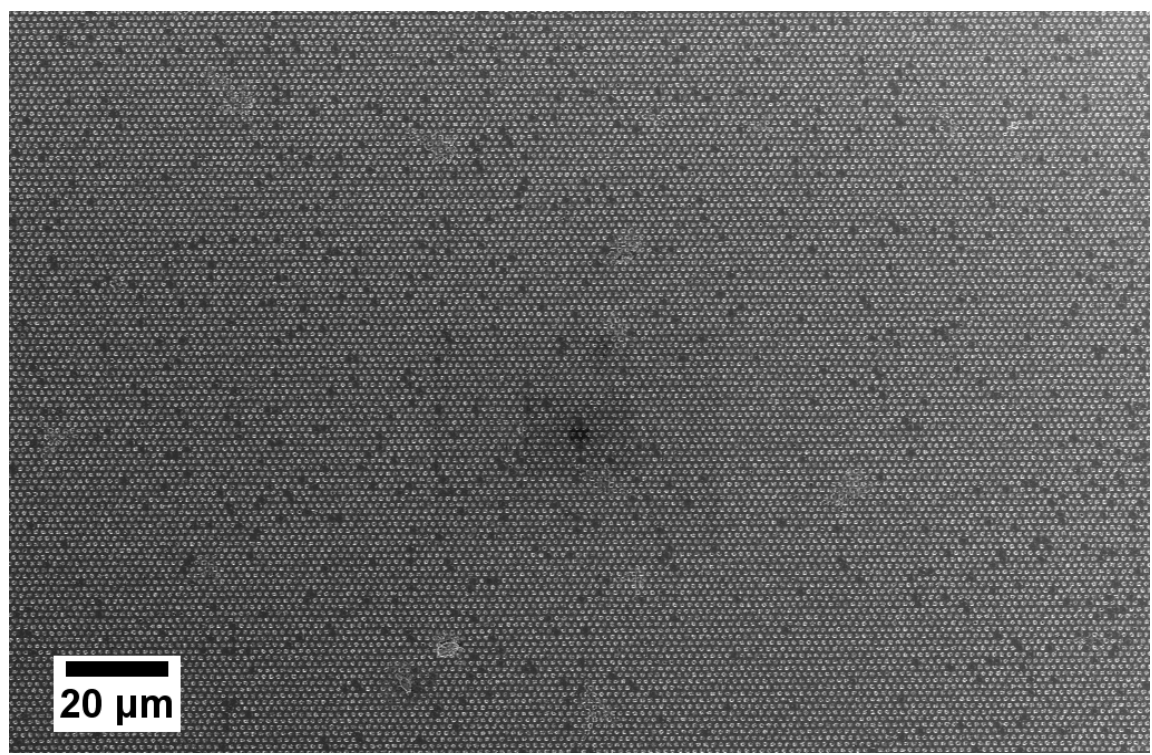

**Figure 4.** Large-field SEM top-view image of an array of PFcMA-CNT hybrid microdots on an ITO substrate obtained by high-temperature capillary stamping.

### 3.2. Pattern quality

On the ITO counterpart substrates arrays of PFcMA-CNT hybrid microdots typically extending 1 $cm^2$ – corresponding to the area of the contact surface of the Ni/$MnO_2$ composite stamps – were deposited (Figure 2c). A large-field SEM image of an array of PFcMA-CNT hybrid microdots is displayed in Figure 4. Inspection of higher-magnified SEM and AFM images (Figure 5) revealed that the nearest-neighbor distance between the PFcMA-CNT hybrid microdots amounted to 1.5 μm, corresponding to the nearest-neighbor distance between the contact elements of the Ni/MnO$_2$ composite stamps. The image field of Figure 4 extends 223 x 146 μm$^2$ and would contain 148 x 112 = 16576 PFcMA-CNT hybrid microdots if the latter were arranged in a defect-free hexagonal array. We evaluated a part of Figure 4 that allowed binarization without contrast saturation comprising an image field of 439 x 592 pixels corresponding to an area of 95 x 129 μm$^2$. This image field would ideally contain ~6237 PFcMA-CNT hybrid microdots. We actually identified 6012 objects with a size of at least 4 pixels in the probed image field using the software ImageJ, that is, 96 % of the objects expected for an ideal lattice. The occurrence of apparent defects may also be related to the surface roughness of the ITO substrates. Closer scrutinization of the SEM image shown as Figure 4 suggests that a certain proportion of the apparent defect sites is actually occupied by PFcMA-CNT hybrid microdots (Figure S4). Unlike most other PFcMA-CNT hybrid microdots, these PDcMA-CNT hybrid microdots do not show prevalent strong edge contrast because they have a modified shape, likely related to the impact of the local substrate topography on the formation of focal contacts with the contact elements of the Ni/MnO$_2$ composite stamps.

It is reasonable to assume the fidelity of the deposited patterns crucially depends on the quality of the used stamps and their ability to form conformal contact with the counterpart substrates as well as on the roughness and the surface chemistry of the counterpart substrates – independent of whether a polymer melt or a polymer solution is used as ink. However, many polymers of interest, such as PFcMA, are only soluble in organic solvents with relatively high vapor pressure. Apart from unwanted structure formation processes including liquid-liquid phase separation[10] and flow processes related to capillarity,[11, 12] even vitrification of the polymer to be stamped may impede stamping. Problems related to solvent evaporation may be overcome by using spongy stamps penetrated by continuous mesopore systems, but this configuration is of limited use if the ink contains components of limited mobility, such as CNTs. Moreover, polymer solutions tend to spread on non-hydrophobized oxidic surfaces with high surface energy. Even if rapid adsorption of non-volatile ink components results in self-inhibition of spreading, only large, flat pancake-like structures are obtained.[31] High-temperature capillary stamping may help overcome these limitations in certain application scenarios.

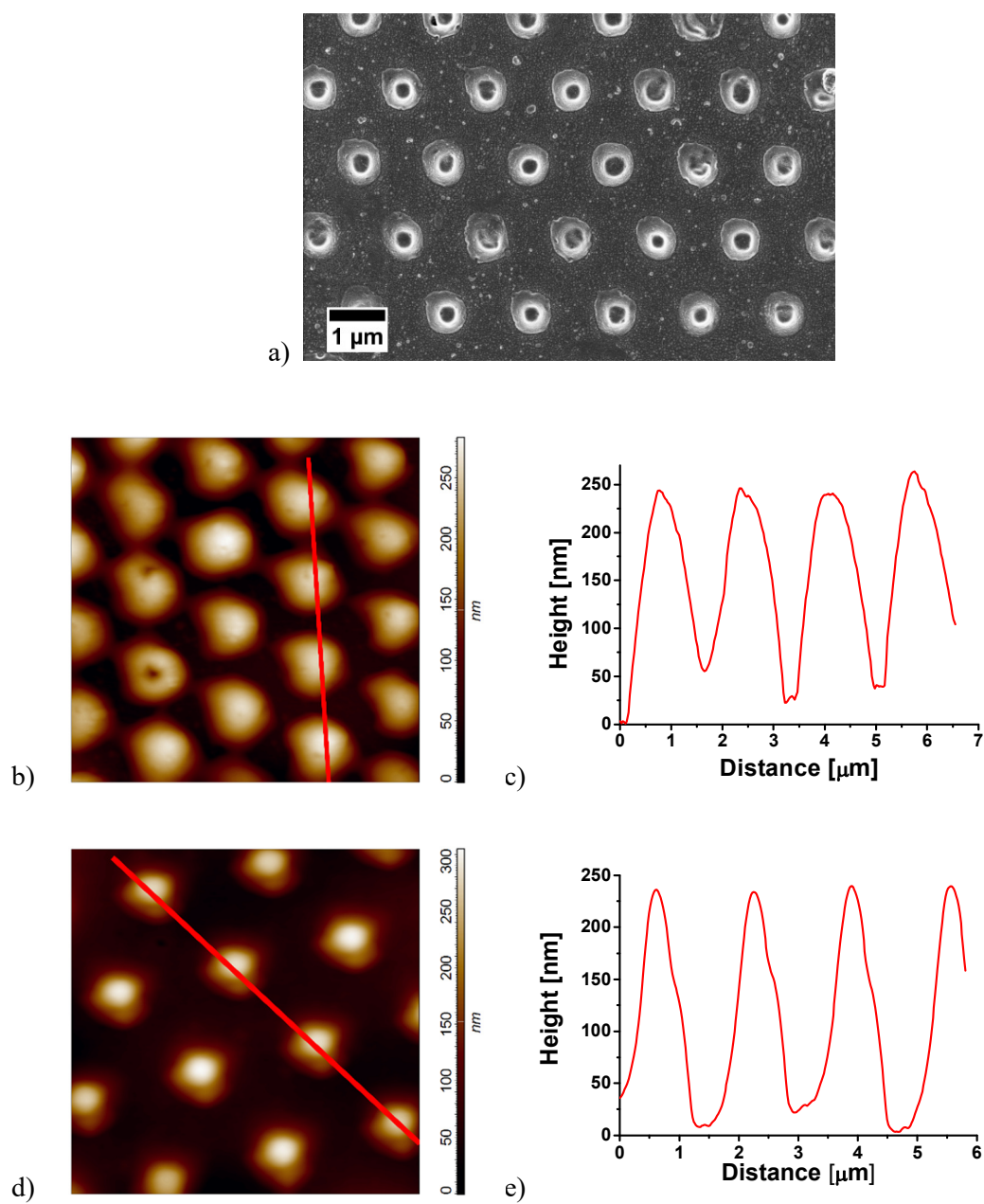

**Figure 5.** Arrays of PFcMA-CNT hybrid microdots on non-modified ITO substrates produced by high-temperature capillary stamping of PFcMA melts containing CNTs. a)-c) First successive stamping step; d), e) fifth successive stamping step without replenishing of ink. a) SEM image. b), d) AFM topography images (the image fields extend b) 8 x 8 μm² and d) 5 x 5μm²); c), e) Topographic height profiles along the red lines in panels b) and d).

## 3.3. Shapes of the PFcMA-CNT hybrid microdots

The diameter of the PFcMA-CNT hybrid microdots amounted to ~800 nm (Figure 5a) and their mean height to 233 nm ± 19 nm (Figure 5b,c, Figure S5). Stamping of pure PFcMA melts without CNTs onto an pristine ITO substrate under exactly the same conditions as described above yielded PFcMA microdots with diameters and heights by and large identical to those of the PFcMA-CNT hybrid microdots; the mean height of the obtained PFcMA microdots amounted to 256 nm ± 37 nm (Figure S6). Also, stamping of PFcMA-CNT mixtures onto ITO substrates modified with PEO-silane but otherwise under the same conditions as described above did not perceptibly alter diameter and height of the PFcMA-CNT hybrid microdots; their mean height amounted to 248 nm ± 30 nm (Figure S7). The molecular weight distribution of PFcMA subjected to high-temperature capillary stamping as described above remained unaltered, as revealed by size exclusion chromatography (Figure S8). After detachment from the ITO counterpart substrates and cooling to room temperature, the Ni/$MnO_2$ composite stamps were still intact (Figure 3d) and could be reused at least five times without replenishment and without apparent changes in the heights of the PFcMA-CNT hybrid microdots. After five consecutive stamping steps we obtained PFcMA-CNT hybrid microdots with a mean height is 239 nm ± 10 nm (Figure 5d, e; Figure S9).

The aliquots of the PFcMA-CNT solution administered by default to load the Ni/$MnO_2$ composite stamps extending 1 $cm^2$ with the PFcMA-CNT mixture contained 10 μg PFcMA and 0.2 μg CNT. We assume that the loading of the Ni/$MnO_2$ composite stamps by dispension of PFcMA-CNT solution did not only result in the deposition of PFcMA-CNT mixture into the mesopores of the $MnO_2$ layer but also into the interstices between the contact elements of the Ni/$MnO_2$ composite stamps (cf. Figures 2 and 3d). We assume that these PFcMA-CNT reservoirs fed contact elements of the Ni/$MnO_2$ composite stamps during high-temperature capillary stamping. The stamping of the PFcMA-CNT mixture onto the ITO surface might also be facilitated by more pronounced van der Waals interactions between PFcMA and ITO as compared to PFcMA and $MoO_2$. The metals contained in ITO, In and Sn, are period 5 elements, whereas Mn is a period 4 element so that the polarizability of ITO is likely be higher than that of $MnO_2$.

The comparative evaluation of both PFcMA-CNT hybrid microdot arrays and PFcMA microdot arrays suggests that the microdot heights are only marginally influenced by the substrate chemistry (unmodified ITO vs. ITO functionalized with PEO-silane) and the manual retraction of the Ni/$MnO_2$ composite stamps. However, loading the Ni/$MnO_2$ composite stamps with significantly reduced amounts of PFcMA per stamp area prior to high-temperature capillary stamping modified the stamped patterns. For example, we reduced the PFcMA concentration in the solution used to load the Ni/$MnO_2$ composite stamps with PFcMA-CNT mixture to one fifth of the default value, i.e., to 2 mg PFcMA per 10 mL solution, while all other variables were kept constant. Subsequent high-temperature capillary stamping onto a non-modified ITO substrate yielded arrays of irregularly shaped doughnut-like PFcMA-CNT hybrid microstructures (Figure S10). This outcome suggests that a certain excess of PFcMA-CNT mixture per stamp area located on the outer $MnO_2$ surface of the $Ni/MnO_2$ composite stamps or in the interstices between the contact elements is needed for reproducible high-temperature capillary stamping.

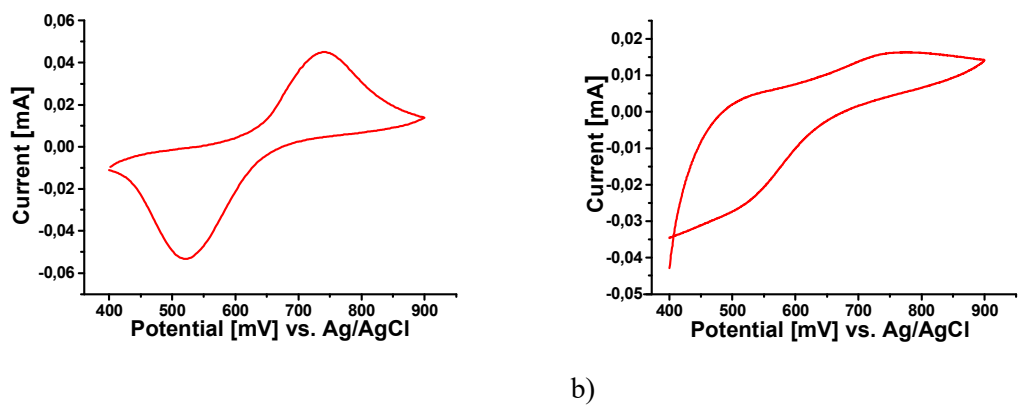

a) b)

**Figure 6**. Cyclic voltammetry of a) PFcMA-CNT hybrid microdot arrays and b) arrays of pure PFcMA microdots deposited on non-modified ITO substrates extending 1 $cm^2$ by high-temperature capillary stamping.

### 3.4. Electrochemical and wettability switching

The impact of the CNTs on the electrochemically induced oxidation and reduction of the PFcMA in PFcMA-CNT hybrid microdots and microdots consisting of pure PFcMA stamped onto pristine ITO substrates was evaluated by cyclic voltammetry. PFcMA-CNT hybrid microdots display narrow and pronounced oxidation and reduction peaks with a peak separation of 220 mV, a maximum oxidation current of 47 µA and a reduction current of 53 µA (Figure 6a). Pure PFcMA microdots also undergo electrochemical oxidation and reduction (Figure 6b). However, the oxidation and reduction peaks are less pronounced because of the lower conductivity of the PFcMA microdots in the absence of CNTs. The apparently larger peak separation can thus hardly be quantified. The oxidation current amounted to 17 µA. Therefore, the presence of CNTs in the PFcMA microdots improved their electrochemical response.

We investigated the wettability of pristine ITO substrates and ITO substrates surface-modified with PEO-silane, which had been patterned with PFcMA-CNT hybrid microdots. To evaluate the impact of the electrochemical switching of the PFcMA-CNT hybrid microdots on the WCAs, we performed five successive electrochemical switching cycles from the reduced state to the oxidized state and *vice versa*. The apparent WCA measured here does not correspond to the equilibrium WCA or to the Young WCA.[43] However, we assume that it reasonably represents the practical wettability of PFcMA-CNT hybrid microdot arrays under operating conditions. The WCA for PFcMA-CNT hybrid microdot arrays on pristine ITO substrates in the reduced state amounted to 112° ± 1° and in the oxidized state to 94° ± 2° (Figure 7a-c). Deposition of the PFcMA-CNT hybrid microdots onto ITO substrates modified with PEO-silane resulted in the reduction of the WCAs by nearly 50°. Thus, a WCA of 65° ± 1° was obtained for the reduced state and of 46° ± 1° for the oxidized state (Figure 7d). X-ray photoelectron spectroscopy (XPS) indicated that the PEO-silane coating in the ITO substrate was still intact after five electrochemical oxidation-reduction cycles (Figure S11). For a continuous PFcMA film spin-coated on an ITO substrate a WCA of 103° ± 1° for the reduced and of 75° ± 3° for the oxidized state was obtained (Figure 7e). In the course of the five successive electrochemical switching cycles no significant systematic change in the WCA values was apparent.

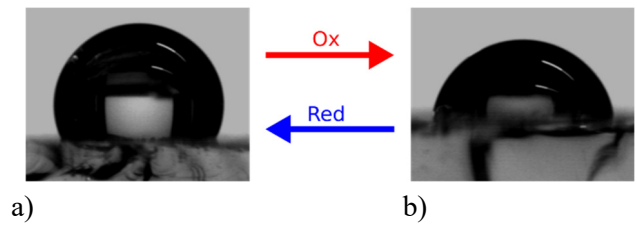

a)  b)

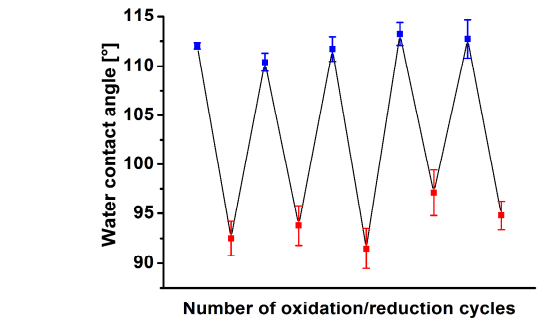

c)

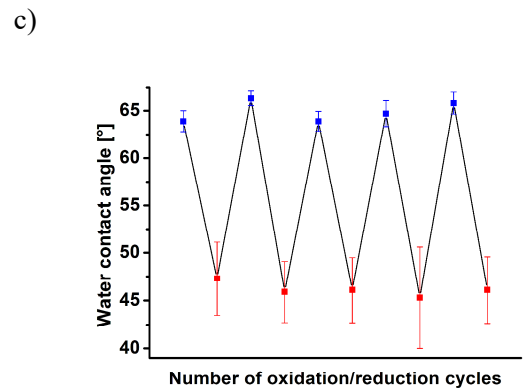

d)

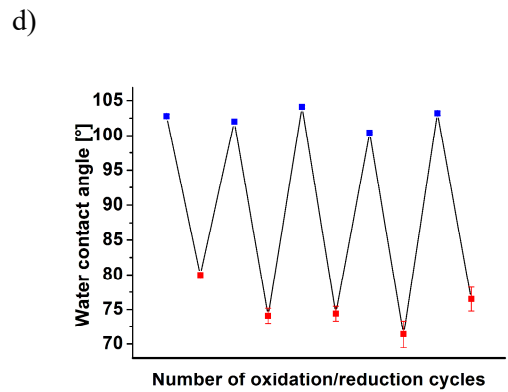

e)

**Figure 7.** a) Electrochemical wettability switching of ITO substrates functionalized with PFcMA-CNT hybrid microdot arrays. Blue denotes the reduced and red the oxidized state. a), b) Sessile water droplets on pristine ITO substrates functionalized with PFcMA-CNT hybrid microdot arrays. The PFcMA is a) in the reduced and b) in the oxidized state. The image widths correspond to 3 mm. c)-d) WCAs of sessile water droplets on c) a pristine ITO substrate patterned with a PFcMA-CNT hybrid microdot array, d) on an ITO substrate modified with PEO-silane followed by patterning with PFcMA-CNT hybrid microdot arrays and e) a continuous PFcMA film with a thickness of 300 nm spin-coated on an ITO substrate.

## 4. Conclusions

We have reported solvent-free high-temperature capillary stamping with Ni/MnO$_2$ composite stamps to generate arrays of stimuli-responsive polymer microdots. In contrast to continuous coatings, crack propagation and delamination are not possible if arrays of discrete microdots are deposited. Moreover, the exposed substrate areas between the microdots can be orthogonally functionalized, resulting in either synergistic or complementary interplay of the properties of the stimuli-responsive polymer and the second functional component. Solvent-free high-temperature capillary stamping of melts prevents problems associated with the use of polymeric solutions containing volatile organic solvents, such as hardly controllable structure formation processes including phase separation induced by solvent evaporation, capillarity-driven flow processes and ink spreading. The method reported here may also enable the lithographic deposition of polymers, which are hardly soluble or which are only soluble in solvents incompatible with commonly used stamps. As example, we generated arrays of hybrid microdots consisting of the stimuli-responsive polymer PFcMA and CNTs by solvent-free high-temperature capillary stamping at elevated temperatures where the PFcMA is molten. PFcMA-CNT hybrids are known to show electrochemically switchable wettability. For solvent-free high-temperature capillary stamping we used specifically designed composite stamps with a topographically patterned contact surface consisting of a nickel core and a MnO$_2$ coating. The latter takes up the PFcMA melt containing the CNTs. The short transport lengths associated with this stamp design allow stamping inks containing additives impeding flow, such as CNTs. Orthogonal functionalization of the ITO substrates between the PFcMA-CNT hybrid microdots with suitable silanes shifted the electrochemical switching range of the water contact angle by nearly 50°.

ASSOCIATED CONTENT

**Supporting Information**. NMR spectrum of as-synthesized PFcMA, SEM images of topographically structured Ni layers, photos of the stamping device, details of Figure 4, AFM images and height histograms of microdots, SEC characterization of PFcMA, XPS spectra of modified ITO substrates. This material is available free of charge via the Internet at http://pubs.acs.org.

AUTHOR INFORMATION

**Corresponding Author**
* Martin Steinhart, Institut für Chemie neuer Materialien and CellNanOs, Universität Osnabrück, Barbarastr. 7, 49076 Osnabrück, Germany; https://orcid.org/0000-0002-5241-8498; Email: martin.steinhart@uos.de

**Author Contributions**
The manuscript was written through contributions of all authors. All authors have given approval to the final version of the manuscript.